\documentclass[twocolumn,twoside,slac]{revtex4}
\ifx\pdftexversion\undefined
  \usepackage[dvips]{graphicx}
\else
  \usepackage[pdftex]{graphicx}
\fi

\usepackage{fancyhdr}
\usepackage{xspace}

\begin{document}

\newcommand{\cwAsk}{\textsc{ask}\xspace}
\newcommand{\cwAskCap}{\textsc{Ask}\xspace}
\newcommand{\cwAtcom}{\textsc{atcom}\xspace}
\newcommand{\cwAtcomCap}{\textsc{Atcom}\xspace}
\newcommand{\cwAthena}{\textsc{athena}\xspace}
\newcommand{\cwAthenaCap}{\textsc{Athena}\xspace}
\newcommand{\cwAtlas}{Atlas\xspace}
\newcommand{\cwAtlasCap}{Atlas\xspace}
\newcommand{\cwAtlfast}{\textsc{atlfast}\xspace}
\newcommand{\cwAtlfastCap}{\textsc{Atlfast}\xspace}
\newcommand{\cwBoost}{\textsc{boost}\xspace}
\newcommand{\cwBoostCap}{\textsc{Boost}\xspace}
\newcommand{\cwBrunel}{\textsc{brunel}\xspace}
\newcommand{\cwBrunelCap}{\textsc{Brunel}\xspace}
\newcommand{\cwCMT}{\textsc{cmt}\xspace}
\newcommand{\cwCMTCap}{\textsc{Cmt}\xspace}
\newcommand{\cwDaVinci}{\textsc{davinci}\xspace}
\newcommand{\cwDaVinciCap}{\textsc{Davinci}\xspace}
\newcommand{\cwDial}{\textsc{dial}\xspace}
\newcommand{\cwDialCap}{\textsc{Dial}\xspace}
\newcommand{\cwDirac}{\textsc{dirac}\xspace}
\newcommand{\cwDiracCap}{\textsc{Dirac}\xspace}
\newcommand{\cwGanga}{\textsc{ganga}\xspace}
\newcommand{\cwGangaCap}{\textsc{Ganga}\xspace}
\newcommand{\cwGaudi}{\textsc{gaudi}\xspace}
\newcommand{\cwGaudiCap}{\textsc{Gaudi}\xspace}
\newcommand{\cwGaudiPython}{\textsc{gaudipython}\xspace}
\newcommand{\cwGaudiPythonCap}{\textsc{Gaudipython}\xspace}
\newcommand{\cwGrid}{grid\xspace}
\newcommand{\cwGridCap}{Grid\xspace}
\newcommand{\cwJoe}{\textsc{joe}\xspace}
\newcommand{\cwJoeCap}{\textsc{Joe}\xspace}
\newcommand{\cwLHCb}{LHCb\xspace}
\newcommand{\cwLHCbCap}{LHCb\xspace}
\newcommand{\cwPybus}{\textsc{pybus}\xspace}
\newcommand{\cwPybusCap}{\textsc{Pybus}\xspace}
\newcommand{\cwPycrust}{\textsc{pycrust}\xspace}
\newcommand{\cwPycrustCap}{\textsc{Pycrust}\xspace}
\newcommand{\cwPyroot}{\textsc{pyroot}\xspace}
\newcommand{\cwPyrootCap}{\textsc{Pyroot}\xspace}
\newcommand{\cwPython}{python\xspace}
\newcommand{\cwPythonCap}{Python\xspace}
\newcommand{\cwRoot}{\textsc{root}\xspace}
\newcommand{\cwRootCap}{\textsc{Root}\xspace}
\newcommand{\cwwxPython}{\textsc{wxPython}\xspace}
\newcommand{\cwwxPythonCap}{\textsc{wxPython}\xspace}
\newcommand{\cwwxWindows}{\textsc{wxWindows}\xspace}
\newcommand{\cwwxWindowsCap}{\textsc{wxWindows}\xspace}

\newcommand{\cwRefFigure}{Fig.$\!$}
\newcommand{\cwRefFigureCap}{Figure}
\newcommand{\cwRefSection}{Section}
\newcommand{\cwRefSectionCap}{Section}

\newcommand{\cwcf}{c.f.\xspace}
\newcommand{\cweg}{e.g.\xspace}
\newcommand{\cwetc}{etc.\xspace}
\newcommand{\cwie}{i.e.\xspace}
\newcommand{\cwndf}{n.d.f.\xspace}
\newcommand{\cwrms}{r.m.s.\xspace}
\newcommand{\cwvs}{v.s.\xspace}
\newcommand{\cwwrt}{w.r.t.\xspace}

\title{GANGA: a user-Grid interface for \cwAtlas and \cwLHCb}

%

%
\author{K.~Harrison}
\affiliation{Cavendish Laboratory, University of Cambridge, CB3 0HE, UK}
\author{W.~T.~L.~P.~Lavrijsen, C.~E.~Tull}
\affiliation{LBNL, Berkeley, CA 94720, USA}
\author{P.~Mato}
\affiliation{CERN, CH-1211 Geneva 23, Switzerland}
\author{A.~Soroko}
\affiliation{Department of Physics, University of Oxford, OX1 3RH, UK}
\author{C.~L.~Tan}
\affiliation{School of Physics and Astronomy, University of Birmingham,
B15 2TT, UK}
\author{N.~Brook}
\affiliation{H.H.~Wills Physics Laboratory, University of Bristol,
BS8 1TL, UK}
\author{R.~W.~L.~Jones}
\affiliation{Department of Physics, University of Lancaster, LA1 4YB, UK}

\begin{abstract}
The \cwGaudiCap/\cwAthenaCap and \cwGridCap Alliance~(\cwGanga) is a
front-end for the configuration, submission, monitoring, bookkeeping, output
collection, and reporting of computing jobs run on a local batch
system or on the \cwGrid.
In particular, \cwGanga handles jobs that use applications written for the
\cwGaudi
software framework shared by the \cwAtlas and \cwLHCb experiments.
\cwGangaCap exploits the commonality of \cwGaudi-based computing jobs, while
insulating against \cwGrid-, batch- and framework-specific technicalities,
to maximize end-user productivity in defining, configuring, and executing jobs.
Designed for a \cwPython-based component architecture, \cwGanga has a modular
underpinning and is therefore well placed for contributing to, and benefiting
from, work in related projects.
Its functionality is accessible both from a scriptable command-line interface,
for expert users and automated tasks, and through a graphical interface,
which simplifies the interaction with \cwGanga for beginning and casual users.

This paper presents the \cwGanga design and implementation, the development of
the underlying software bus architecture, and the functionality
of the first public \cwGanga release.
\end{abstract}

\maketitle

\thispagestyle{fancy}

\section{INTRODUCTION}

The \cwAtlasCap~\cite{ref:Atlas94} and \cwLHCb~\cite{ref:LHCb98} collaborations
will perform physics studies at the high-energy, high-luminosity, Large Hadron
Collider~(LHC)~\cite{ref:LHC95}, scheduled to start operation at CERN
in 2007.
The data volumes produced by each experiment are in the petabyte range,
and will be processed with computing resources that are distributed over
national centers, regional centers, and individual institutes.
To fully exploit these distributed facilities, and to allow the participating
physicists to share resources in a coordinated manner, use will be made of
\cwGrid services.

The \cwGaudi/\cwAthena~\cite{ref:gaudi,ref:athena} software
framework\footnote{From here on referred to as the ``\cwGaudi framework'' or
simply ``\cwGaudi.''} used by \cwLHCb and \cwAtlas, is designed to
support all event-processing applications, including simulation, reconstruction,
and physics analysis.
A joint project has been set up to develop a front-end that will aid in the
handling of framework-based jobs, and performs the tasks necessary to run these
jobs either locally or on the \cwGrid.
This front-end is the \cwGaudiCap/\cwAthenaCap and \cwGridCap Alliance, or
\cwGanga~\cite{ref:ganga}.

\cwGangaCap covers all phases of a job's life cycle: creation, configuration,
splitting and recollection, script generation, file transfer to and from worker
nodes, submission, run-time setup, monitoring, and reporting.
In the specific case of \cwGaudi jobs, the job configuration also includes
selection of the algorithms to be run, definition of algorithm properties, and
specification of inputs and outputs.  \cwGangaCap relies on middleware from
other projects, such as Globus~\cite{ref:globus} and EDG~\cite{ref:edg}, to
perform Grid-based operations, but makes use of the middleware functionality
transparent to the \cwGanga user.

In this paper, we present the \cwGanga design and the choices made for its
implementation.
We report on the work done in developing the software bus that is a key feature
of the design, and we describe the functionality of the current \cwGanga
release.

\section{\label{sec:design}GANGA DESIGN}

\subsection{Overview}

\begin{figure*}[t]
\begin{center}
\includegraphics[width=0.8\textwidth,angle=270]{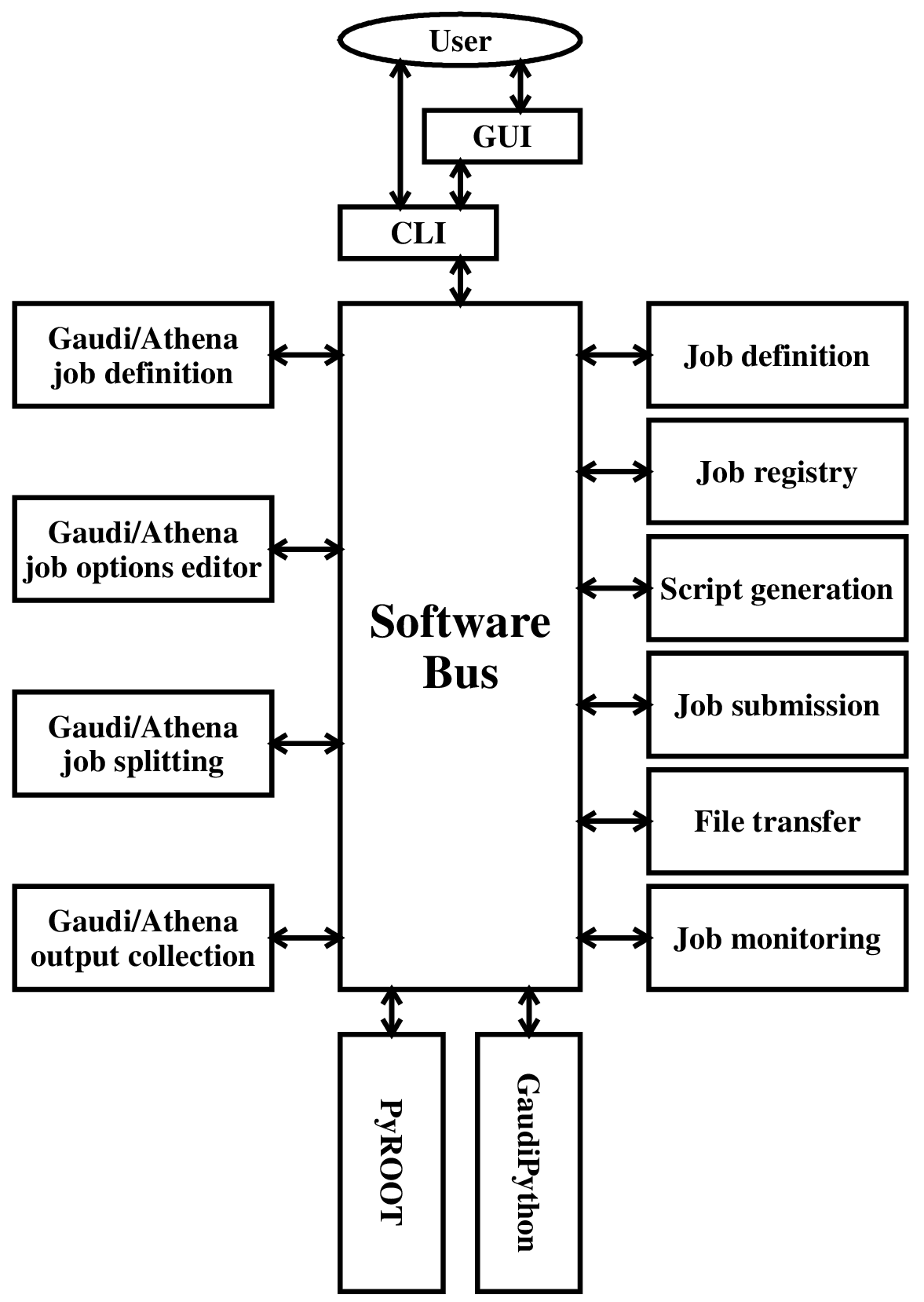}
\caption{Schematic representation of the \cwGanga design, which is based on
components interacting via a software bus.
The user issues commands either via the graphical user interface (GUI) or via
the command-line interface (CLI).
\cwGangaCap components of general applicability are shown on the right side of
the software bus, whereas \cwGanga components dedicated to specific requirements
of the \cwGaudi framework are shown on the left.
Components external to \cwGanga are shown at the bottom: \cwGaudiPython and
\cwPyroot are \cwPython interfaces to \cwGaudi and \cwRoot respectively.}
\label{fig:architecture}
\end{center}
\end{figure*}

\cwGangaCap is being implemented in \cwPython~\cite{ref:python}, an interpreted
scripting language, using an object-oriented approach, and following a component
architecture.
The \cwPython programming language is simple to use, supports object-oriented
programming, and is portable.  By virtue of the possibilities it allows for
extending and embedding features, \cwPython is also effective as a
software ``glue.''
A standard \cwPython installation comes with an interactive interpreter, an
Integrated Development Environment (IDE), and a large set of ready-to-use
modules.
The implementation of the component architecture underlying the \cwGanga
design benefits greatly from \cwPython's support for modular software
development.
The components of \cwGanga interact with one another through, and are
managed by, a so-called ``software bus''~\cite{ref:mato03a}, of which a
prototype implementation is described in \cwRefSection~\ref{sec:pybus}.
This interplay is displayed graphically in \cwRefFigure~\ref{fig:architecture}.

As considered here, a component is a unit of software that can be connected to,
or detached from, the overall system, and brings with it a discrete,
well-defined, and circumscribed functionality.
In practical terms, it is a \cwPython module (either written in \cwPython or
embedded) that follows a few non-intrusive conventions.
The component-based approach has the advantages that it allows two or more
developers to work in parallel on well-separated tasks, and that it allows
reuse of components from other systems that have a similar architecture.
The initial functionality of the software bus is provided by the \cwPython
interpreter itself.
In particular, the interpreter allows for the dynamic loading of modules, after
which the bus can use introspection to bind method calls dynamically, and
to manage
components throughout their life cycle.
The functionality of the \cwGanga components can be accessed through
a Command-Line Interface (CLI), and through a Graphical User Interface
(GUI), built on a common Application-Programmer Interface (API).  All
actions performed by the user through the GUI can be invoked through
the CLI, allowing capture of a GUI session in an editable CLI script.

The components can be divided among three categories: general, application
specific, and external.
Further developments will add components as needed.
With reference to \cwRefFigure~\ref{fig:architecture}, they are discussed below.

\subsection{\label{sec:generalcomponents}Generally applicable components}

Although the first priority is to deal with \cwGaudi jobs, \cwGanga has a
set of core components suitable for job-handling tasks in a wide range
of application areas.
These core components provide implementations for the job definition,
the editing
of job options, the splitting of jobs based on user provided configuration and
job templates, and the output collection.
In \cwRefFigure~\ref{fig:architecture}, these components are shown on the
right side of the software bus.

The job definition characterizes a \cwGanga job in terms of the following:

\begin{itemize}

\item
The name chosen as the job's identifier.

\item
The workflow (see below) indicating the operations to be performed when the job
is run.

\item
The computing resources required for the job to run to completion.

\item
The job status (in preparation, submitted, running, completed).

\end{itemize}

A job workflow is represented as a sequence of elements (executables,
parameters, input/output files, and so on), with the action to be performed by,
and on, each element implicitly defined.
Resources required to run a job, for example minimum CPU time or minimum memory
size, are specified as a list of attribute-value pairs, using a syntax not tied
to any particular computing system.
The job-definition component implements a job class, and classes corresponding
to various workflow elements.

Other \cwGanga components of general applicability performing operations on,
for, or using job objects:

\begin{itemize}

\item
A job-registry component allows for the storage and recovery of information for
job objects, and allows for job objects to be serialized.

\item
A script-generation component translates a job's workflow into the set of
(\cwPython) instructions to be executed when the job is run.

\item
A job-submission component takes care of submitting the workflow script to a
destination indicated by the user, creating Job Description Language (JDL) files
where necessary, and translating the resource requests into the format expected
by the target system (European DataGrid (EDG), GridPP \cwGrid, US-ATLAS testbed,
local PBS queue, and so on).

\item
A file-transfer component handles transfers between sites of job input and
output files, this usually involving the addition of appropriate commands to the
workflow script at the time of job submission.

\item
A job monitoring component keeps track of job status, and allows for
user-initiated and scheduled status queries.

\end{itemize}

\subsection{User group specific components}

\cwGangaCap can be optimized for a given user group, through the addition of
application-specific components.
For the current user groups, \cwAtlas and \cwLHCb, specialized
components exist that incorporate knowledge of the \cwGaudi framework:

\begin{itemize}

\item
A component for \cwGaudi job definition adds classes for workflow elements not
dealt with by the general-purpose job definition component.
For example, applications based on \cwGaudi are packaged using a
configuration management tool (\cwCMT)~\cite{ref:cmt}, which requires its own
workflow elements.
Also, this component provides workflow templates covering a variety of common
tasks, such as simulating events and running an analysis on some dataset.

\item
A component for \cwGaudi job-options editing allows selection of the algorithms
to be run and modification of algorithm properties.

\item
A component for \cwGaudi job splitting allows for large jobs to be broken down
into smaller sub-jobs, for example by examining the list of input data files and
creating jobs for subsets of these files.

\item
A component for \cwGaudi output collection merges outputs from sub-jobs where
this is possible, for example when the output files contain data sets like
histograms and/or ntuples.

\end{itemize}

Specialized components for other application areas are readily added.  The
subdivision into general and specialized components, and the grouping
together of specialized components dedicated to a particular problem domain,
allows new user groups to identify quickly the components that match their
needs, and will improve \cwGanga stability.

\subsection{External components}

The functionality of the components developed specifically for \cwGanga is
supplemented by the functionality of external components.
These include all modules of the \cwPython standard library, and also
non-\cwPython components for which an appropriate interface has been written.
Two components of note in the latter category, both interfaced using
\cwBoost~\cite{ref:boost}, allow access to the services of the \cwGaudi framework
itself, and to the full functionality of the \cwRoot analysis
framework~\cite{ref:root}.

\section{\label{sec:pybus}PYTHON SOFTWARE BUS}

\subsection{Functionality}

To first order, the software bus functionality required by \cwGanga is provided
by the \cwPython interpreter itself.
The main features not offered by the interpreter, but nevertheless desirable
are:

\begin{itemize}

\item{\bf Symbolic component names}\\
\cwPythonCap modules are loaded on the basis of their names, which are mapped
one-to-one onto names in the file system, sometimes including (part of)
the directory structure.
Components should, instead, be loaded on the basis of the functionality that
they promise.

\item{\bf Replacing a connected component}\\\
This is different from the standard reload functionality, which loads a new
version of a current module and doesn't rebind any outstanding references.
A component, however, may need to be completely replaced by another component,
meaning that the latter must be reloaded deep, and references into the old
component must be replaced by equivalent references into the new component,
wherever possible.

\item{\bf Disconnecting components}\\
A standard \cwPython module does not get unloaded until all outstanding
references disappear.
This is common behavior in many off-the-shelf component architectures.
However, it should be possible to propagate the unloading of a component through
the whole system, allowing for more natural interactive use. 

\item{\bf Configuration and dependencies}\\
Since \cwPython modules simply execute \cwPython code, their configuration and
dependencies are usually resolved locally.
Components should be able to advertise their configurable parameters and their
dependencies, such that it is also possible to manage the configuration
externally and/or globally.

\end{itemize}

The software bus should also support a User Interface Presentation Layer (UIPL),
through which the configuration, input and output parameters, and functionality
of components can be connected to user interface elements.
The bus inspects the component for presentable elements, including (if
applicable) their type, range, name, and documentation.
It subsequently requests the user interface to supply elements that are capable
of providing a display of and/or interaction with each of the parameters, based
on their type, range, \cwetc
Both the interface element and the component should then be hooked through the
UIPL.

For example, assume that a configuration parameter of a component is of a
boolean type.
This parameter can then map onto \cweg\ a checkbox in a GUI.
The bus requests the GUI to provide a display of the boolean value, gets a
checkbox in return, and it subscribes the checkbox to a value holder in the
UIPL.
It also subscribes itself to the value holder.
Changes by the checkbox, changes the value in the value holder, which in turn
causes a notification to the bus, which sets the value in the component.

Mapping through an UIPL has the advantages that simple user interfaces can be
created automatically, and more sophisticated user interfaces can be relatively
easily peeled off and replaced, since they never access the actual underlying
component directly.

\subsection{The PyBus prototype}

A prototype of a software bus (\cwPybus) has been written to explore the
possibilities for implementing the above features in a user-level \cwPython
module, rather than in a framework.
That is, \cwPybus is a client of the \cwPython interpreter and does not have any
privileges over other modules.
This means that components written for \cwPybus will act as components when used
in conjunction with the bus, or as \cwPython modules when used without.
Conversely, \cwPython modules that were not written as \cwPybus components can
still be loaded and managed by the software bus, assuming that they adhere to
standard \cwPython conventions concerning privacy and dependencies.  

When a user connects a component to \cwPybus in order to make it available to
the system, he or she can load it using the logical, functional, or actual
name.
Components that are available to \cwPybus must be registered under logical
names, optionally advertising under functional names the public contracts that
they fulfill (their ``interfaces,'' but note that \cwPython does not explicitly
support interfaces in the sense of Java or C++).
If a component is not registered, it can only be connected using its actual name,
which is the name that would be used in the standard way of identifying a
\cwPython module.
Unlike the actual name, which has to be unique, the logical name and functional
names may be claimed by more than one component.
\cwPybusCap will choose among the available components on the basis of its own
configuration, a priority scheme, or a direct action from the user. 

In the process of connecting a module, \cwPybus will look for any conventional
parameters (starting with ``PyBus\_'') in the dictionary of the module that
contains the component.
These parameters may describe dependencies, new components to be registered,
post-connect or pre-disconnect configuration, and so on.
It is purely optional for a module to provide any of these parameters and
\cwPybus will use some heuristics if they are absent.  For example, all public
identifiers are considered part of the interface, so that any module can be
connected as if it were a component.
The user can decide the name under which the module should be connected, which
can be any alias that stands in for the name to be used in the user code,
following the `as' option for standard \cwPython imports.
If no alias is provided, the logical name is used.
\cwPybusCap keeps track of these aliases. 

\cwPybusCap allows components to be replaced.
In order to do this, it uses the garbage collection and reference-counting
information of the \cwPython interpreter to track down any outstanding
references, and acts accordingly.  Some references, for example those to
variables or instances, are rebound, whereas others, for example object
instances, are destroyed.
Disconnecting a component is rather similar to replacing it, with the only
exception that no references are rebound: all are destroyed. 

\cwPythonCap allows user modules to intercept the importing of other modules, by
replacing the import hook.
This mechanism allows the \cwPybus implementation to bookmark those modules that
are imported during the process of connecting a component, and thus manage
component dependencies.  When disconnecting a component, the modules that it
loaded are not automatically removed, since the interpreter itself holds on to a
reference to them, even after all user-level modules are unloaded.
There is no real reason to force the unloading of standard modules, but if the
modules are components that are connected to \cwPybus, their unloading is
mandatory.
When a new component is loaded, which in turn loads other components, then
\cwPybus needs to resolve the lookup of those components anew: these dependent
components may have changed, or different components may be chosen this time
around, for a variety of reasons. 

The implementation of the prototype of a software bus has been mostly successful.
The are a few issues left for embedded components, but for pure \cwPython
components it has been shown that it is possible to implement the component
architecture features missing in the \cwPython interpreter with a user-level
\cwPython module.

\section{THE FIRST RELEASE}

\subsection{Overview of functionality}

The \cwGanga design, including possibilities for creating, saving, modifying,
submitting and monitoring jobs, has been partly implemented and released.
The tools implemented should be suitable for a wide range of tasks, but we have
initially focused on running one type of job for \cwAtlas and one type of job
for \cwLHCb, focusing on the \cwAtlfast~\cite{ref:atlfast} fast simulation in
the case
of the former, and the \cwDaVinci~\cite{ref:davinci} analysis in the case of
the latter.
Optimization for other types of applications essentially means creating more
templates, which is a straightforward procedure.
The current release allows jobs to be submitted to the European DataGrid, and
to local PBS and LSF queues.
Communication between components in the prototype is via the \cwPython
interpreter, with the sophistication of \cwPybus to be added later.

The first release does not yet fulfill all requirements, but it already helps
the user to perform a number of tasks that otherwise would have to be done
manually.
For example, the creation of JDL files necessary for job submission to
the EDG Testbed, and the generation of scripts to submit jobs to other batch
systems, has been automated.

Most parameters relevant for \cwGaudi-based jobs have reasonable default
values, so that
a user only has to supply minimal information to create and configure
a new job.
Existing jobs can easily be copied, renamed, and deleted.
When a job is created, it is presented as a template that can be edited by
the user.
After making the proper modifications, the user can submit the job for execution
with a single command.
If the generated scripts and job-options files need to be verified before
submission,
the user can perform the job set up with the configure command.
Job splitting is achieved through loading and executing a splitter script,
with support for both default and user scripts.

When a job is submitted, \cwGanga starts to monitor the job state by
periodically
querying the appropriate computing system.
This process can be stopped or started manually at any time.

When a job is running, it can be killed, if so desired.
On job completion, the output is automatically transferred to a dedicated output
directory or to any other location described in the job
output files.

Below, we give details of the \cwGanga's current job-handling capabilities
and the main graphics features: the GUI and the job-options editor.

\subsection{Core, application and job handlers}

\begin{figure*}[t]
\begin{center}
\includegraphics[width=0.8\textwidth,angle=0]{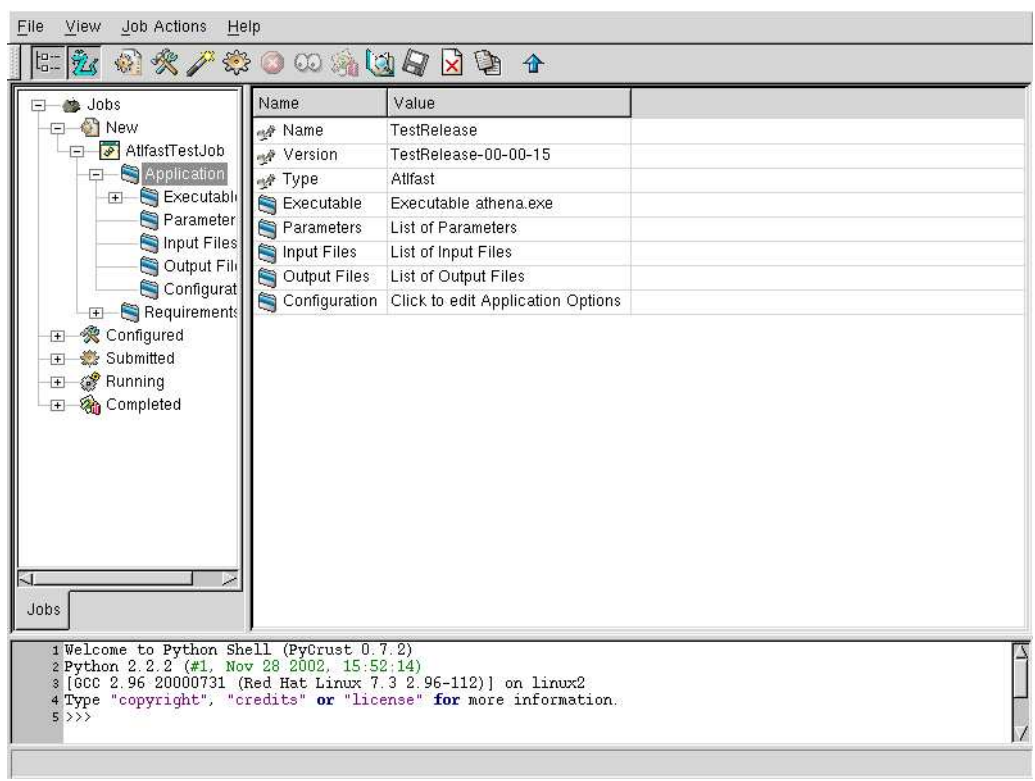}
\caption{Screenshot showing the general layout of the \cwGanga GUI.}
\label{fig:gui}
\end{center}
\end{figure*}

The job registry (described in \ref{sec:generalcomponents}) keeps records, in
the form of metadata, about all user jobs.
It also allows operations on jobs, such as job creation,
configuration, submission, termination, and provides the job monitoring service.
For job serialization, the registry uses a job catalogue, which in turn
maintains the information about all saved jobs \cwetc

Each user job is represented by an instance, which contains information
about the job state,
and references to the requirements, application and handler objects.
The specific steps required for job configuration, submission, and
monitoring, which differ for each computing system, are delegated to a
job handler.

In order to complete a step, the handler uses the requirements, which are
common by design, and it adds its own attributes that are relevant for the
particular computing system.
Examples of common requirements are minimal size of physical memory or minimal
size of available disk space, examples of specific attributes are queue time
limits and bookkeeping accounts.
For job monitoring, the handler provides information about the job state that
is system dependent.
\cwGanga currently has components containing job handlers to work
with the local computer, a local PBS or LSF batch system, or the EDG.

In the \cwGanga design, a distinction is made between ``jobs'' and
``applications,'' in order to have the possibility to run the same application
on different computing systems.
An application in \cwGanga terminology represents a computer program that
the user wants to execute (the executable), together with any necessary
configuration parameters, required input, and expected output files.
The executable is specified by image location, name, and version.
Application parameters, which may be files, include a type description with the
actual value.
The input and output files are described by their name and (desired) locations.
Methods are available to get them from their storage location, and to transfer
files to and from worker nodes, tailored to each of the supported computing
systems.
In \cwGanga there is currently support for the local system copy
command, the gridftp transfer protocol, and the EDG sandboxes mechanism.
The transfer method is set up automatically by the job handler during the
configuration of a job.
An interface to the EDG Replica Catalogue to translate logical file names is
also implemented, and future \cwGanga releases will contain more advanced data
management tools to work with the \cwGrid.

Like jobs, applications plug into the appropriate application handlers based on
the type of application.
Currently, \cwGanga offers a generic application handler, which can
be used with types of application, but with the disadvantage that it
provides little help with configuration; a \cwGaudi handler, for use with
general applications written for the \cwGaudi framework; and two handlers
specific to \cwDaVinci and \cwAtlfast.

\subsection{Graphical user interface}

\begin{figure*}[t]
\begin{center}
\includegraphics[width=0.8\textwidth,angle=0]{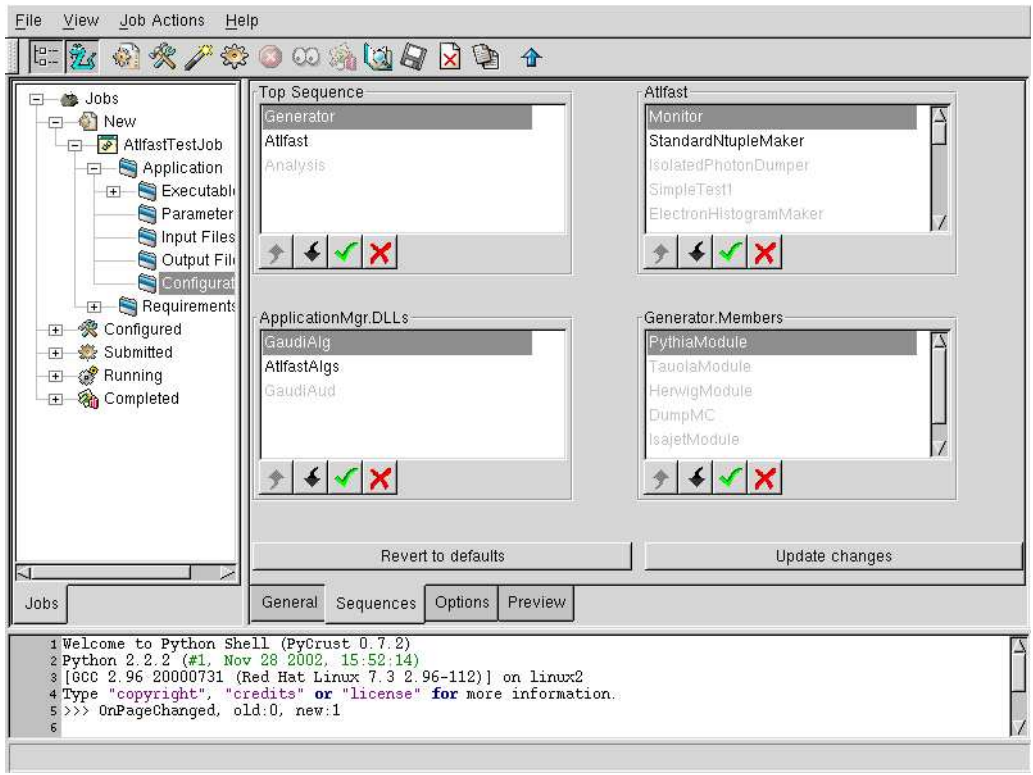}
\caption{Screenshot of \cwGanga, showing one of the windows presented by the
job options editor.
The example window is for defining sequences.}
\label{fig:joe}
\end{center}
\end{figure*}

The GUI, like the rest of Ganga, is implemented in \cwPython, and makes
use of \cwwxPython, the
extension module that embeds the \cwwxWindows platform independent application
framework.
\cwwxWindowsCap is implemented in C++, and is a layer on top of the native
operating and windowing system.
The design of the GUI is based on a mapping of major \cwGanga core classes --
jobs, applications, files, and so on --
onto the corresponding GUI classes.
These classes (GUI handlers) add the hooks necessary to provide interaction
with the graphical elements of the interface, on the top of the functionality of
the underlying core classes.
The GUI component also includes some \cwGanga specific extensions of standard
\cwwxWindows widgets (dialogues, panels, list and tree controls).

The basic display of the \cwGanga GUI is shown in
\cwRefFigure~\ref{fig:gui}.
The layout of the window consists of three main parts: there is a tree
control on the left that displays job folders, which themselves may be
folders, in their respective states.
There is a multi-purpose panel on the right, which facilitates many displays
(see also \cwRefFigure~\ref{fig:joe}), and which in \cwRefFigure~\ref{fig:gui}
consists of a list control.
Finally, there is an embedded \cwPython shell (\cwPycrust, itself designed
for use with \cwwxPython).

With the advanced (expert) view option on, the job folders, in the tree of job
states, display a hierarchy of all job-related values and parameters.
The most important values are brought to the top of the tree, less important
ones are hidden in the branches.
The normal (user) view stops at the level of jobs and gives access to the most
important parameters only.
The user can also choose to hide the tree control completely.
The list control displays the content of the selected folder in the job
tree.
With a double mouse click it is possible to edit most of the values shown in
this control list.
The lower part of the frame is reserved for the \cwPython shell, which doubles
as a log window.
The shell does not only allow the execution of any \cwPython command, but it
also permits access to, and modification of, any GUI widget.
The shell, too, can be hidden if desired.

Actions on jobs can be performed through a menu, using a toolbar,
or via pop-up menus called by a right click in various locations.

When the monitoring service is switched on, jobs move automatically from one
folder to another as their states evolve.
To avoid delays in the program response to the user input, the
monitoring service runs on its own thread.
It posts customized events whenever the state of a job changes and the display
is to be updated.
For GUI updates not related to job monitoring, \cwGanga handles the standard
update events posted by the \cwwxWindows framework.

\subsection{Job-options editor}

The job-options editor (\cwJoe) has been developed in the context of
the work on \cwAtlfast, and allows the user to customize \cwAtlfast jobs from
within the \cwGanga environment.

The main features of \cwJoe are summarized as follows:

\begin{itemize}

\item
\cwJoeCap, through its GUI (\cwRefFigure~\ref{fig:joe}), assists the user in
customizing \cwAtlfast job options from within \cwGanga.
This helps to eliminate human errors arising from incorrectly spelt
options/values and incorrect syntax.

The user can define sequences/lists (\cweg\ TopSequence, Generator.Members,
ApplicationMgr.DLLs, \cwetc) by enabling/disabling entries and arranging them
in any desired sequence.
Currently, there are no restrictions placed on these user-defined sequences for
them to be meaningful.

\cwJoeCap incorporates an option-type aware user interface presentation selector
that essentially chooses the correct presentation format at run-time
(\cweg\ drop-down menus for discrete choice options, arbitrary value entry for
simple choice options, value appending for list-type options, \cwetc) for
individual job options based on their attributes/characteristics.

\item
Job option settings for commonly performed jobs may be saved and reused.
This saves the user the work of re-entering option values for subsequent
jobs, especially if only minor modifications are needed.
With default values for options built-in, the user is able to revert to the
default settings when required.

\item
Once all the options have been set, the preview function allows the user to
check that the created script is as required.

\item
In accordance with the basic \cwGanga requirements, all functionality of the
editor is available on the \cwPython command line without the GUI (not without
a certain degree of visual inconvenience of course).
Users and developers alike can make use of this API.

\end{itemize}

Since the current version of \cwJoe is very basic, improvements are in the
pipeline:

\begin{itemize}

\item
The editor is to be decoupled from the \cwAtlfast application handler and made
available to be used in a generic \cwGaudi environment.
A generic editor can be used to perform job-option customization of full
simulation, reconstruction and analysis jobs.
To make this possible, the editor's dependence on hard-coded \cwPython data
structures must be removed and replaced with a database that can be queried.

\item
Option attributes that enable \cwJoe to dynamically choose appropriate
presentation formats for individual job options are currently hard-coded into
the data structure referred to above.
Future versions will attempt to make deductions about job option attributes at
run-time.

\item
With the foreseen decline in the use of text-based options files, the editor
will support the creation of \cwPython scripts instead.

\item
Editable previews of options files, to allow the user to make last minute
changes.

\item
The move from \cwAtlfast to full simulation and reconstruction will see at least
a tenfold increase in the number of configurable job options.
It will not be useful to display all options indiscriminately; some form of
information hiding is required.

The ``Favorite-options first'' feature will further speed up the user's task of
job-options modifications by placing frequently used options at the top of the
list and perhaps hide the not so frequently used ones.

\item
Although rudimentary option-value checks are performed, the more important
range checking is not yet available.
This feature requires permitted ranges (\cwie\ sensible values) of individual
options to be attributes of individual options.

\end{itemize}

\cwJoeCap showcases the extensibility of the \cwGanga user interface.
Future extensions can be developed and incorporated in the same way.

\section{OUTLOOK}
The first release of the \cwGanga package has been made available, and user
feedback is being collected from both \cwAtlas and \cwLHCb collaborators.
The development schedule laid out for the remainder of calendar year 2003 is
targeted at providing a product to satisfy requirements for the \cwAtlas and
\cwLHCb data challenges.
Cooperation and integration with existing projects (\cwAsk~\cite{ref:ask},
\cwAtcom~\cite{ref:atcom}, \cwDirac~\cite{ref:dirac},
\cwDial~\cite{ref:dial}) is foreseen in order to meet in time these
requirements.
The \cwGanga project will then continue to keep pace with and adapt to the
ever-evolving \cwGrid middleware services.

\begin{acknowledgments}
This work was partly supported by the Office of Science.
High Energy Physics, U.S. Department of Energy under Contract
No.~DE-AC03-76SF00098.
\end{acknowledgments}


\end{document}